\documentclass{appolb}
\usepackage{graphicx}

\begin{document}
\title{The ELENA project at CERN
\thanks{Presented at II SYmposium on Applied Nuclear Physics and Innovative Technologies in Krak\'{o}w}%
}
\author{W. Oelert$^{a}$\\representing the ELENA collaboration
\address{Johannes Gutenberg-Universit\"at Institut f\"ur Physik,  \\  Staudingerweg 7, D - 55128 Mainz, Germany}
}
\maketitle
\begin{abstract}
CERN has a longstanding tradition of pursuing fundamental physics on extreme low and high energy scales. The present physics knowledge is successfully described by the Standard Model and the General Relativity. In the anti-matter regime many predictions of this established theory still remain experimentally unverified and one of the most fundamental open problems in physics concerns the question of asymmetry between particles: why is the observable and visible universe apparently composed almost entirely of matter and not of anti-matter?
There is a huge interest in the very compelling scientific case for anti-hydrogen and low energy anti-proton physics, here to name especially the Workshop on “New Opportunities in the Physics Landscape at CERN” which was convened in May 2009 by the CERN Directorate and culminated in the decision for the final approval of the construction of the Extra Low ENergy Antiproton (ELENA) ring by the Research Board in June 2011. ELENA is a CERN project aiming to construct a small 30 m circumference synchrotron to further decelerate anti-protons from the Antiproton Decelerator (AD) from 5.3 MeV down to 100 keV. 
\end{abstract}
\PACS{PACS numbers come here}
  \section{INTRODUCTION}
\begin{figure}[h]
\center
\includegraphics[width=0.99\textwidth]{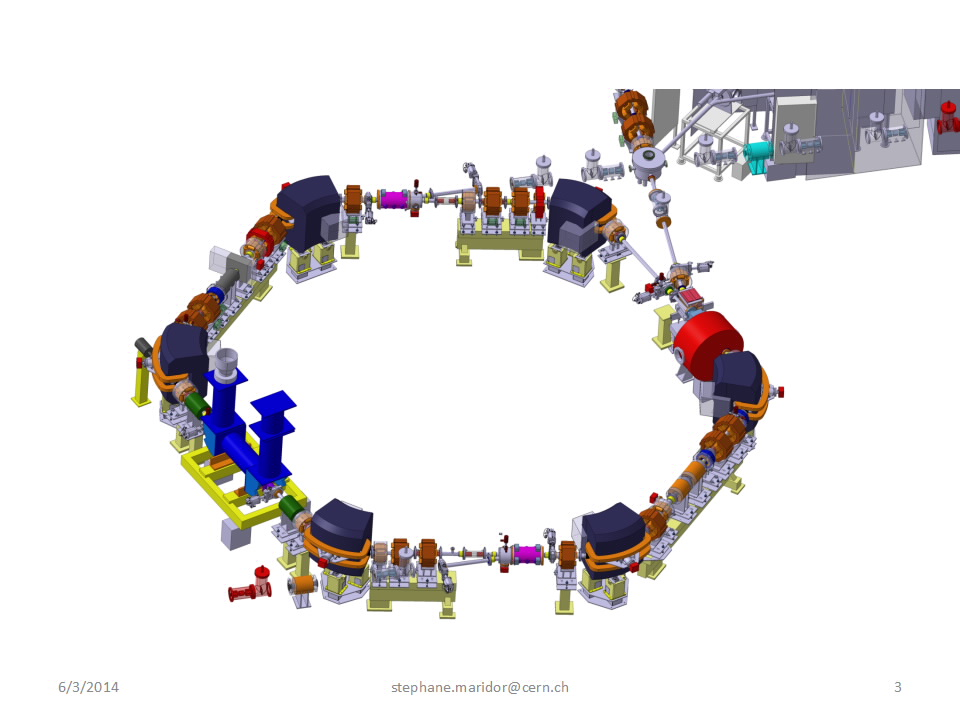}
\caption{Layout of the ELENA ring and its components.}   
\end{figure}
The layout of the ELENA ring~\cite{OEL26, CHOH14} with its main  
components is sketched in Figure 1.
Controlled deceleration in a synchrotron equipped with an electron
cooler to reduce the emittances in all three planes will allow the 
existing AD experiments to increase substantially their anti-proton 
capture efficiencies and render new experiments possible. The ELENA 
design is now well advanced and the project is moving to the 
implementation phase. Component design and construction are taking 
place at present for installation foreseen during the second half of 
2015 and beginning of 2016 followed by ring commissioning until the 
end of 2016. New electrostatic transfer lines to the experiments will 
be installed and commissioned during the first half of 2017 followed 
by the first physics operation with ELENA. Basic limitations like 
intra beam scattering limiting the emittances obtained under electron 
cooling and direct space charge effects is reviewed and the status of 
the project is reported. In that sense ELENA is an upgrade of the 
Anti-proton Decelerator (AD)~\cite{Belo} at CERN and is devoted to 
experiments for physics using low energy anti-protons. 
The AD is a unique facility constructed after the completion 
of the exploitation of the Low Energy Antiproton Ring LEAR and 
providing low 5.3 MeV energy antiprotons to experiments.  Most 
experiments further decelerate the beam using degrader foils or a 
decelerating Radio Frequency Quadrupole RFQD and then capture the 
beam in traps.  Both processes to decelerate are not optimal and lead 
to significant antiproton losses.  Deceleration with a degrader foil 
is limited by energy straggling such that, even with optimized 
thickness, many antiprotons are stopped in the foil and annihilate 
there and many still have a too high energy to be trapped; this 
results in a trapping efficiency well below 1\%.  Matching to the 
RFQD is difficult, in particular in the longitudinal plane, and 
physical emittances increase during the deceleration resulting in 
losses.  \\
The ELENA project aims at constructing a small 30.4 m circumference 
synchrotron to improve the trapping efficiencies of existing 
experiments by one to two orders of magnitude by controlled 
deceleration in a small synchrotron and reduction of the emittances 
with an electron  cooler.   New  types  of  experiments  will  become 
feasible.  The antiprotons will be injected at 5.3 MeV, an energy 
reachable safely in the AD and then 
decelerated down to 100 keV possible with such a small ring. Electron 
cooling will be applied at an intermediate plateau and at the final 
energy.  Moreover, ELENA will not send the full available intensity 
in one bunch to one experiment, but send several (baseline four) 
bunches with lower intensity to several experiments.  The resulting 
longer runs for the experiments are considered an advantage despite 
the lower intensity. A sketch of the ELENA machine the AD hall is
shown in Figure 2.
\begin{figure}[h]
\center
\includegraphics[width=0.99\textwidth]{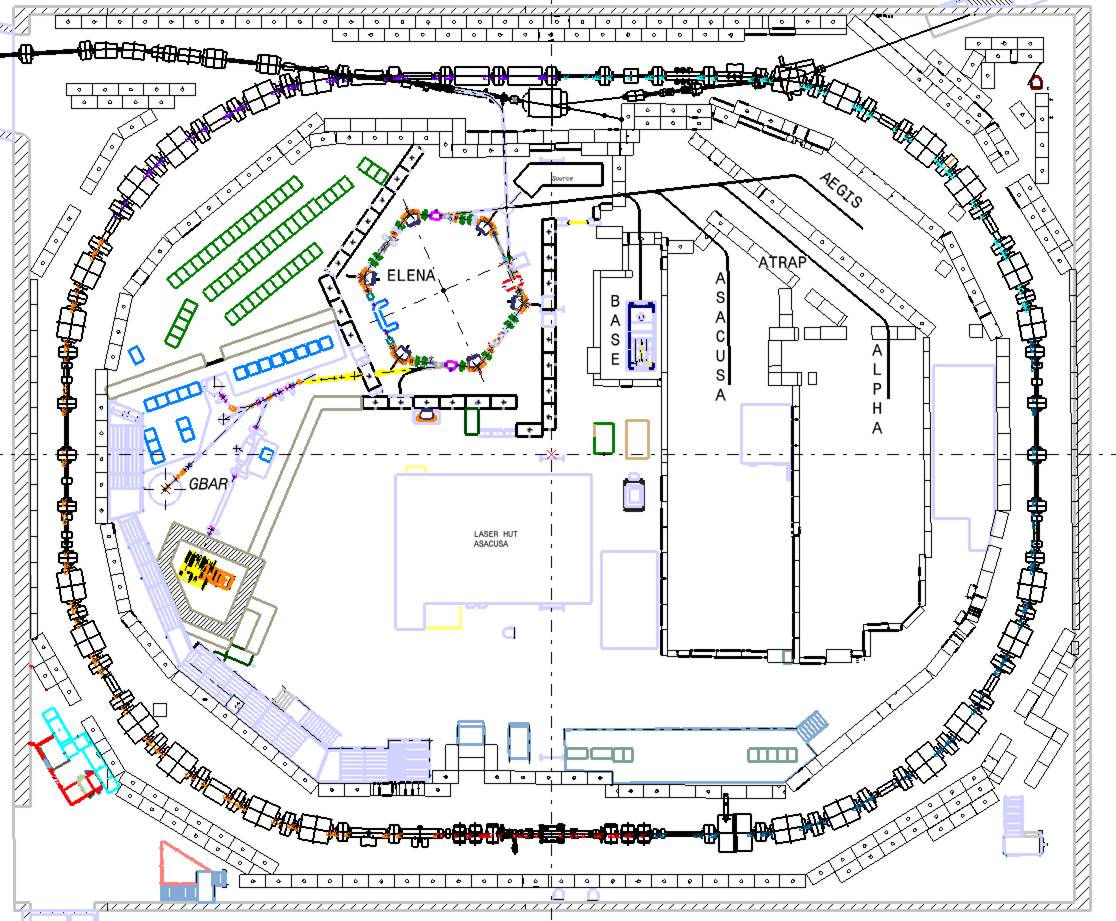}
\caption{ELENA in the AD Hall, where in the inner part of the AD ring the experiments are hosted as well.}      
\end{figure}
The main physics topics at ELENA are the anti-hydrogen production 
with consecutive studies of the features of this anti-matter atom, 
the anti-proton nucleon interaction by testing the QED to extremely
high precision as well as the demanding question of the gravitation force between matter and anti-matter.

\section{ELENA main features}
The main features and possible issues of the new facility ELENA are:
\begin{itemize}
\item
ELENA is operated at an unusually low energy for a synchrotron with a 
magnetic focusing structure.  Thus, any possible performance 
limitation has to be evaluated with particular attention to the low 
beam energy.  Many of the features listed below are the consequence 
of this unusual energy range.
\item
The machine will be located inside the existing AD hall. 
This is an economic solution as no large additional building is needed to house the new ring and experiments
and further more allows for keeping existing experiments
at their present location. A smaller new building has been
completed recently in order to free space in the AD hall
for the ELENA ring and a second experimental area.
\item
The lattice design has to cope with typical difficulties
for small machines as few quadrupole families to adjust
optics parameters, 
constraints on lattice parameters to be fulfilled and to
deal with strong focusing due to the bending magnets.  
An important condition was to find a layout suitable for
installation in the AD hall and compatible with the
position of the injection and the two extractions 
towards the foreseen experimental areas.  
The ELENA ring has hexagonal shape and two-fold periodicity
(neglecting the perturbation of the lattice due to the 
cooler).  Two slightly longer straight 
sections without quadrupoles house the electron cooler with 
associated equipment and the injection line. Three
quadrupole families (one magnet of each family in each 
of the remaining four sections) allow adjusting the lattice.
\item
A good magnetic field quality has to be guaranteed despite the very 
low magnetic fields required and remanence and hysteresis effect.  
From the beginning of the project, it had been foreseen to apply 
“thinning”, i.e. mixing of non-magnetic stainless steel laminations 
with magnetic laminations, for the main bending magnets; this 
increases the magnetic flux density in the magnetic laminations and 
reduces hysteresis effects.  Quadrupole prototypes are constructed to 
test whether “thinning” is appropriate for quadrupoles and, possibly, 
sextupoles as well. Orbit correctors will be constructed without 
magnetic cores to avoid any effects related to hysteresis.
\item
Electron cooling will be applied at an intermediate energy of around 
650 keV to reduce emittances before further deceleration and avoid 
losses, and at the final energy 100 keV.  The design of the ELENA 
cooler is based on the one constructed for the L-LSR ring in 
Kyoto, but with parameters optimised for our case.  First simulations 
of electron cooling at 100 keV predicted final energy spreads of the 
coasting beam by about one order of magnitude larger than expected 
initially.  With adiabatic bunching without electron cooling, this 
would have led to energy spreads at the limit of the acceptance of 
the transfer lines and inacceptable for some experiments.  In order 
to reduce the energy spread of bunches sent to the experiment, 
bunched beam electron cooling will be applied by keeping the cooler 
switched on during the capture process.
\item
Emittance blow-up due to intra beam scattering is expected to 
be the main performance limitation and to determine, together with 
the performance of electron cooling, the characteristics of the beams 
extracted and sent to the experiment.
\item
Beam diagnostics is challenging due to the low intensity and 
velocity.  For example, the lowest beam currents are well below 1 A, 
which is well below the capabilities of standard slow beam current 
transformers.  Thus, the intensity of coasting beams during electron 
cooling will be determined by Schottky diagnostics using optimised 
pick-ups.  Further diagnostics in the ring are low noise position 
pick-ups (precision better than 0.1 mm), a tune measurement BBQ 
system, a scraper to determine destructively emittances and possibly 
a ionization profile monitor (feasibility under study).  Diagnostics 
in the lines comprises TV stations, GEM monitors and “micro-wire” 
profile monitors.

\item
Cross sections for interactions with rest gas molecules as scattering 
out of the acceptance or emittance blow-up become large at low 
energies.  These effects have to be evaluated with care paying 
attention to the very low energy in order not to overestimate 
emittance blow-up.  The machine will be fully bakeable and 
equipped with NEG coatings wherever possible in order to reach the 
challenging nominal pressure of 3 10-12 Torr and to guarantee that 
interactions with rest gas is not the main performance limitation.
\item
An RF system with a rather modest RF voltage of less than 500 V is 
sufficient for deceleration and to create short bunches at 
extraction.  However, the system has to cover a large dynamic range.
\item
Direct space charge detuning is a significant effect despite the low 
intensity due the low energy and short bunches required for the 
experiment and would result in tune shift of almost -0.4 with only 
one extracted bunch.  As mitigation measure, the available intensity 
will be split into several bunches to serve several experiments 
almost simultaneously.
\item
Extraction and transfer to the experiments is based electrostatic 
elements as this is an efficient low-cost solution at these low 
energies.
\item
Commissioning of the ELENA ring will be done mainly with an external 
source providing H- ions or protons in parallel to AD operation for 
experiments.  This allows injecting beam with a higher repetition 
rate than would be possible with a antiprotons and a bunch every ≈100 
s from the AD and is expected to speed up commissioning despite the 
fact this implies starting at the at lowest energy. 
\end{itemize}

\section{AD-ELENA performance} 
In operation since end of 1999 the AD is an "All-IN-ONE" machine 
which collects anti-protons, decelerates them
 in three steps via first stochastic and then electron cooling down to a momentum 
of 100~MeV/c or 5.3~MeV kinetic energy. 
A typical cycle lasts $\approx$110~seconds,
delivers about 3 $\times 10~^7 ~\bar p$ per pulse of $\sim$150~ns length. 
The foreseen cycle of ELENA has a length of about 25 seconds and
fits well into the AD cycle. Thus,  the overall timing sequence of available
anti-proton bursts will still be determined by the AD cycle.
The basic parameters for AD and for ELENA are presented in Table 1 and 
Table 2, respectively. 
\begin{table}[h]
\caption{AD -- Basic Parameters} 
$~~~~~~~$\\  
\begin{tabular}{lll}
\hline\noalign{\smallskip}
Item & Value & Dimension  \\
\noalign{\smallskip}\hline\noalign{\smallskip}
Circumference &182 & m \\
Production beam &1.5 $\times$ 10$^{13}$ & protons/cycle \\
Injected and cooled beam  &4 $\times$ 10$^{7}$ & anti-protons/cycle \\
Beam momenta max/min  &3.6 / 0.1 & GeV/c \\
Momenta for beam cooling$~~~~~~~~~~~~~~~$ \\
$~~~~~~~$stochastic cooling   &3.6 and 2.0 & GeV/c \\
$~~~~~~~$electron cooling     &0.3 and 0.1 & GeV/c \\
Transverse emittances   &200 ~--~ 1 & $\pi$  mm  mrad \\
Momentum spread &6 $\times$ 10$^{-2}~ $ -- 1 $\times$ 10$^{-4}~$     & $\Delta p/p$\\
Average vacuum pressure &4 $\times$ 10$^{-10}$ & Torr \\
Cycle length  &100 & s \\
Deceleration efficiency  &85 & \% \\
\noalign{\smallskip}\hline
\end{tabular}
\end{table}
\\[1.0cm]
\begin{table}[h]
\caption{ELENA -- Basic Parameters}  
$~~~~~~~$\\      
\begin{tabular}{lll}
\hline\noalign{\smallskip}
Item & Value & Dimension  \\
\noalign{\smallskip}\hline\noalign{\smallskip}
Circumference &30.4 & m \\
Beam momenta max/min  &100 / 13.7$~~~~~~~~~~~~~~~~~~$ & MeV/c \\
Energy range max/min  &5.3 /0.1 &MeV \\
Working point         &2.3/1.3   &Q$_x$/Q$_y$ \\
Ring acceptance       &75        &$\pi$ mm mrad \\
Intensity of injected/ejected beam     &3.0 10$^7$ / 1.8 10$^7$ &  \\
Number of extracted bunches  & $\le$4 & \\
Emittances (h/v) of extracted bunches &4/4   &$\pi$ mm mrad [95 \%]\\
$\Delta p /p$ of extracted beam & 2.5 10$^{-3}$ & [95 \%]\\
Bunch length  at 100~keV &1.3 / 300 & m / ns \\
Required (dynamic) vacuum    &3 10$^{-12} $ &Torr  \\
\noalign{\smallskip}\hline
\end{tabular}
\end{table}

\section{Physics at ELENA}

ELENA will increase the number of useful anti-protons by one to two 
orders of magnitude and will allow serving up to four experiments 
simultaneously. Beam lines from ELENA to various experiments are show 
in Figure 3.
\begin{figure*}[h]
\center
\includegraphics[width=0.99\textwidth]{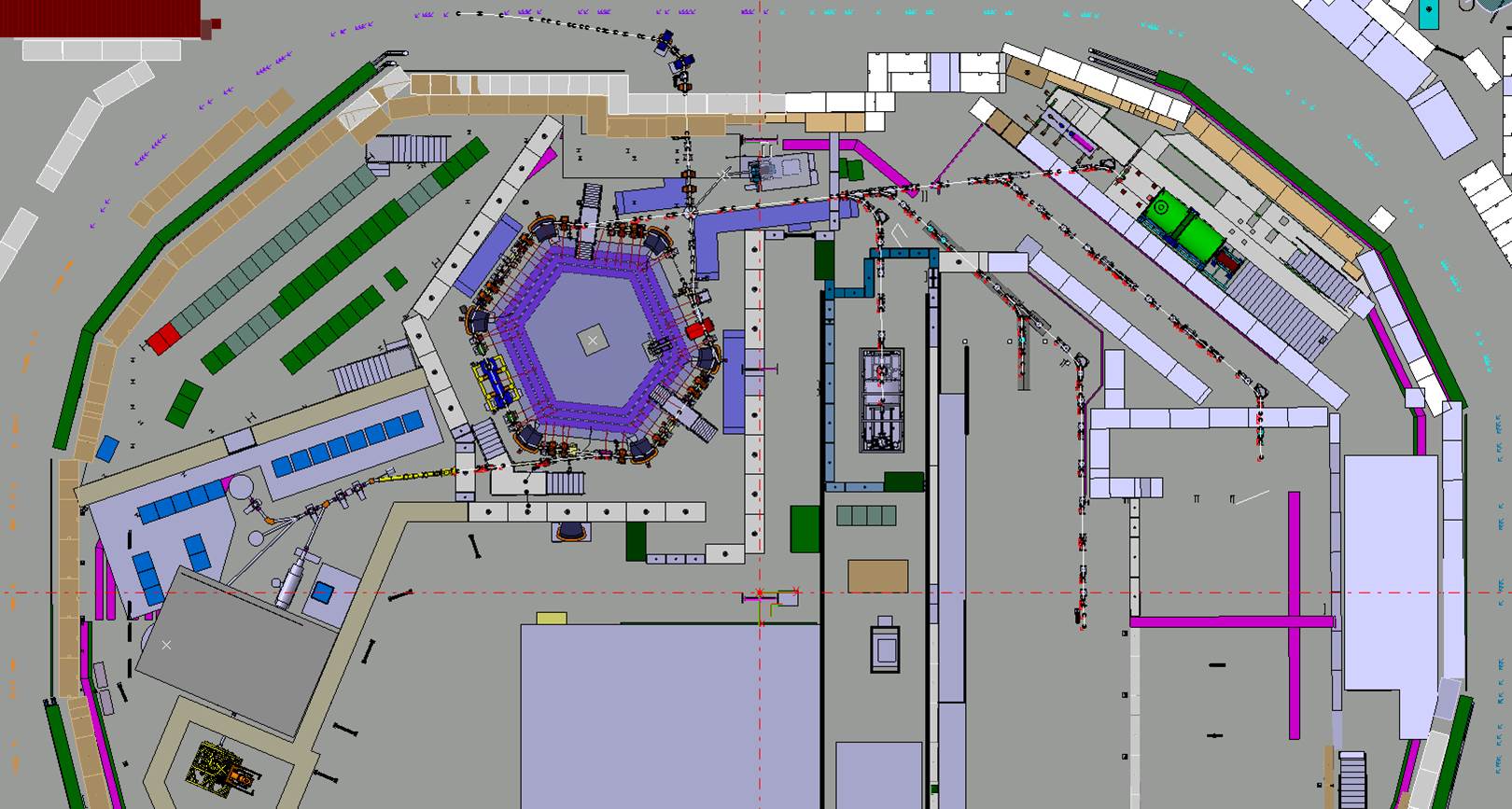}
\caption{ELENA and the beam lines to the different experiments in the AD Hall.}      
\end{figure*}
The motivation for ELENA was driven by the perspective to increase 
the efficiency of the anti-proton facility with its compact and time 
consuming experiments and by the steadily growing interest of 
additional research groups to share the available beam time. In 
particular, concrete motivations from the physics side arise in a 
number of theoretical approaches extending the established model 
frame work including a consistent unified description of the corner 
stones of physics: Lorentz symmetry, quantum mechanics and gravity. 
Experiments with anti-protons will substantially increase the 
knowledge of atomic, nuclear and particle physics by testing 
precisely familiar interactions and fundamental constants, by 
studying discrete symmetries and by searching for new interactions.
These days anti-hydrogen atoms are produced frequently by three 
collaborations at the AD: ATRAP~\cite{ATRAP}, ALPHA~\cite{ALPHA}, and 
ASACUSA~\cite{ASACU} employing essentially similar methods. Whereas 
ATRAP and ALPHA produce anti-hydrogen at rest, ASACUSA produces a 
beam of these atoms for hyperfine transition studies in low magnetic 
fields.
In 2002 both first the ATHENA collaboration and shortly thereafter 
the ATRAP group announced the creation of the first "cold"
anti-hydrogen. Still, since the neutral anti-hydrogen atom is 
unaffected by the electric fields used to trap its charged components 
the anti-hydrogen hits the trap walls and annihilates very soon after 
its creation.

High-precision tests of the properties of anti-hydrogen in magnetic 
minimum traps can only be performed if the anti-hydrogen atoms are 
cold enough to be hold in place for a relatively long time. The 
anti-hydrogen atoms have a magnetic moment which interacts with an 
inhomogeneous magnetic field; low field seeking anti-hydrogen atoms 
can be trapped in a magnetic minimum. In fall 2010 the ALPHA 
collaboration reported first the success of 38 trapped anti-hydrogen 
atoms. A year later life times of more than 15 minutes of the 
anti-hydrogen atoms were observed by both collaborations ALPHA and 
ATRAP. Finally ALPHA reported on the very first spectroscopy of an 
anti-matter atom demonstrating the observation of resonant quantum 
transitions in anti-hydrogen by manipulating the internal spin state.

In addition, new supplementary experiments at the AD, as 
AEGIS~\cite{AEGIS} and GBAR~\cite{GBAR} are presently preparing for 
precise measurements of the gravitational interaction between matter 
and anti-matter. Further the BASE~\cite{BASE} collaboration suggested 
a measurement of the magnetic moment of the anti-proton increasing 
the precision by a factor of 1000 as compared to the successful 
recent result of the ATRAP group at the AD.
 
The basic concept of the ELENA ring and transfer lines to the 
experiment is completed, the TDR~\cite{CHOH14} is published and the 
design and the integration of components are ongoing. The place 
inside the AD hall, where ELENA will be installed, will only be made 
available during the first part of 2015, when the kicker equipment 
can be moved to the new building, which is finished in its body 
shell, see Figure 4. 
\begin{figure*}[h]
\center
\includegraphics[width=0.70\textwidth]{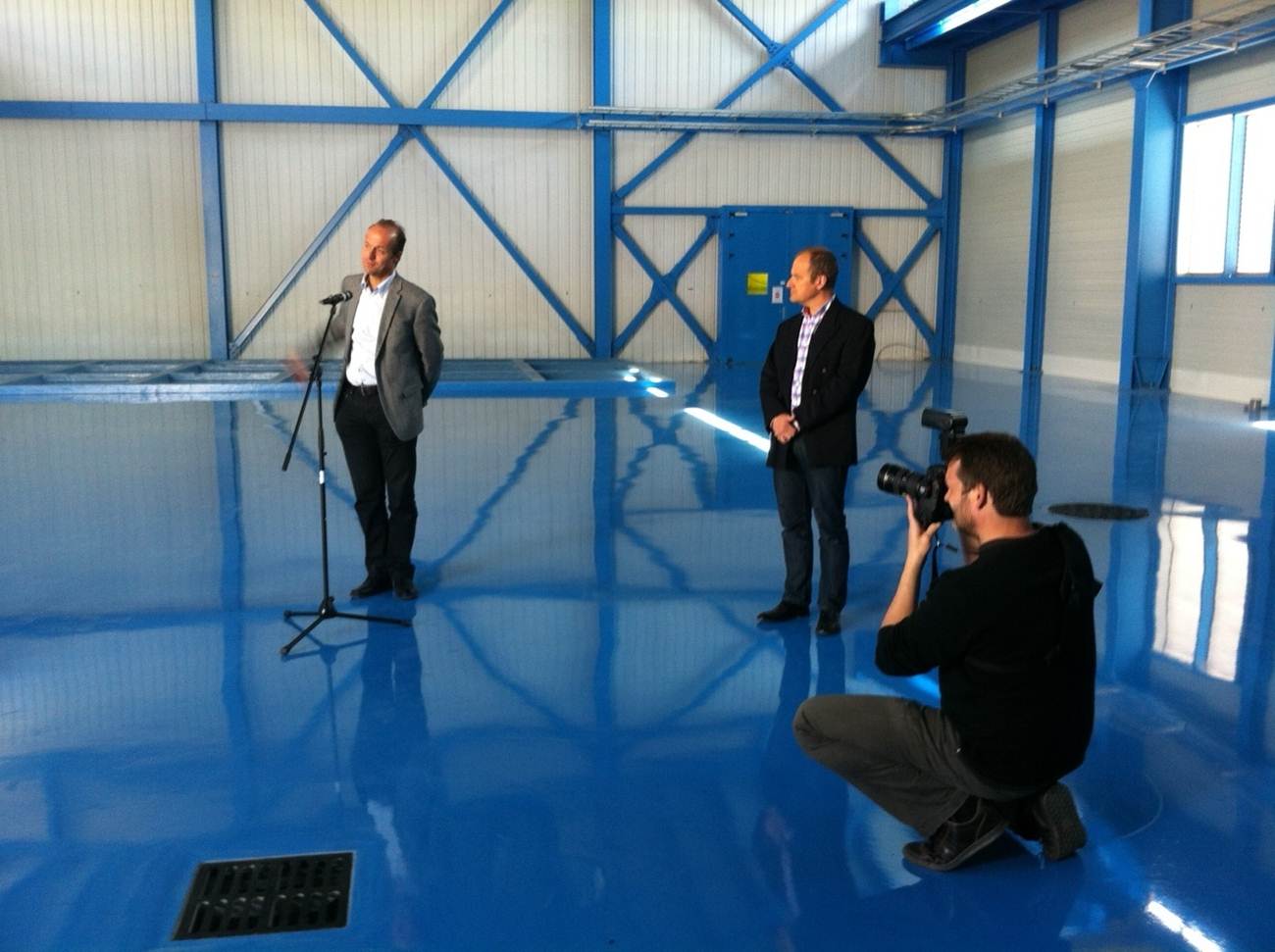}

\caption{Inside the new building 393 during the opening ceremony in April 2014.}      
\end{figure*}
After installation in 2015 and 2016, the commissioning of the ELENA 
ring is planned for the second half of 2016 in parallel to AD 
operation mainly with the help of a dedicated source delivering 100 
keV protons and H- ions. During the first part of 2017, the existing 
magnetic transfer lines from the AD to the experiments will be 
dismantled and the new electrostatic lines from ELENA installed. 
After commissioning of the new electrostatic lines, first beams for 
physics are expected in the second half of 2017.

\end{document}